\documentclass[a4paper,11pt]{article}
\usepackage{pos}
\usepackage{braket}
\usepackage{simplewick}
\usepackage{graphicx}

\title{The Lambda 1405 at the $SU(3)$ point in lattice QCD}
\author*[a]{Javier Suarez Sucunza}
\author[a,b]{Thomas Luu}
\author[a]{Carsten Urbach}

\affiliation[a]{Helmholtz-Institut für Strahlen- und Kernphysik (Theorie) and Bethe Center for Theoretical Physics, \\  Universität Bonn, 53115 Bonn, Germany}

\affiliation[b]{Institute for Advanced Simulation 4 (IAS-4),\\ Forschungszentrum Jülich, Germany}

\emailAdd{suarez@hiskp.uni-bonn.de}
\emailAdd{}

\abstract{The pole structure of the $\Lambda(1405)$ has been a topic of debate for a long time. Chiral perturbation theory predicts that its experimental spectrum may be explained by a two pole structure originating in the $SU(3)$ chiral dynamics of the baryon-meson interaction. The $SU(3)$-symmetric flavor point is readily accessible in lattice QCD, in this work we study the baryon-meson states directly at this point. We construct interpolation operators that belong to the irreducible representations of $SU(3)$ that are attractive in the channel with the quantum numbers of the (singlet and two octets). The extracted energy levels can be used as input for chiral perturbation theory to find the poles associated with each representation. The relevant correlation functions are computed on $SU(3)$-symmetric ensembles with $M_{\pi}\approx 714$ MeV using the distillation technique.}

\FullConference{The 42nd International Symposium on Lattice Field Theory (LATTICE2025)
2-8 November 2025
Tata Institute of Fundamental Research, Mumbai, India% 
}

\begin{document}
\maketitle

\section{Introduction}

	The $\Lambda$(1405) is a resonance found in the $\pi\Sigma$ channel with a mass of 1405 MeV and a with of 50.5 MeV, and it sits just below the  $\overline{K}N$ threshold. It is categorized as a four star resonance in the PDG \cite{PDG} with spin-parity $J^{P}=\frac{1}{2}^{-}$; strangeness $S=-1$; and Isospin  $I=0$. Thus, the expected quark content of this resonance is $(uds)$, however, it is known that the naive quark model is not fully capable of accounting for this resonance \cite{PhysRevD.18.4187}. Most notably, the predicted mass for these quantum numbers is larger than what is actually observed. A potential explanation for the $\Lambda(1405)$ comes from Unitary Chiral Perturbation Theory (UChPT) \cite{Jido:2003cb}. In this framework the $\Lambda$(1405) is explained as a two pole resonance originating in the baryon meson interaction at the flavor $SU(3)$ symmetric point.

At the $SU(3)$ point the Baryon-Meson interaction can be decomposed into irreducible representations
\begin{equation}
  \mathbf{8}_{B}\otimes\mathbf{8}_{M} = \textcolor{blue}{\mathbf{1}}\oplus\textcolor{blue}{\mathbf{8}}\oplus \textcolor{blue}{\mathbf{8'}} \oplus\mathbf{10} \oplus\mathbf{\overline{10}} \oplus \mathbf{27},
\end{equation}
where the baryon octet is that with positive parity. In the channels with the quantum numbers of the $\Lambda$(1405) the singlet and both octet representations are attractive which can lead to bound states. The decuplet  representations are non interacting in these channels and the 27-plet is repulsive. As one moves away from the $SU(3)$ point the bound states will move in the complex energy plane. Two poles, one coming from the singlet the other from one of the octets, will move into the region of the $\Lambda$ (1405) giving way to the two-pole resonance. The $\Lambda$(1405) has become the prime example of a two pole resonance and, therefore, its understanding is an important benchmark in the understanding of the  $SU(3)$ hadron dynamics.

To study the   $\Lambda$(1405) with UChPT, one needs some external input, in order to fix the low energy constants. This input can come in the form of experimental data, lattice QCD energy levels or a combination of both \cite{Pittler:2025upn}. The first studies in lattice QCD into the $\Lambda$(1405) looked at isolating the lowest laying finite volume spectrum using single baryon three quark operators \cite{Gubler2016,Menadue2011,Engel2012,Engel2013,Nemoto2003,Burch2006,Takahashi2009,Meinel2021,Hall2014}. More recently there have been studies using a coupled channel scattering analysis with Baryon-Meson operators in gauge configurations away from the $SU(3)$ limit and at a pion mass of $M_\pi \approx 200$ MeV \cite{BaryonScatteringBaSc:2023zvt,BaryonScatteringBaSc:2023ori}. The results in these papers support the UChPT picture of a two pole resonance. A separate study, using the HAL QCD method has investigated the $S$-wave Meson-Baryon interaction in the singlet and octet channels in the $SU(3)$ limit \cite{Murakami:2023phq,Murakami:2025oig}. For a review on the $\Lambda$(1405) considering theoretical and experimental developments see \cite{Mai:2020ltx}.

Because the origin of the two pole structure is related to the strength of the attractive forces of the decomposed interaction, looking directly at the energy levels of the different irreducible representations can help understand the nature of the $\Lambda$ (1405) and its chiral dynamics. The $SU(3)$ point can be directly accessed with lattice QCD simulations, providing a way to study the origin of the two-pole structure directly at the symmetric point. In UChPT at the leading order, the poles coming from the octets are degenerate in the $SU(3)$ limit. This accidental symmetry is broken once once incorporates next-to-leading order (NLO) terms \cite{Guo:2023wes}. Furthermore the trajectories that the poles follow connecting the $SU(3)$ limit with the physical point vary a lot from. To study these properties, we use Baryon-Meson interpolation operators that belong to the attractive irreducible representations to measure their corresponding energy levels. We look at the location of these energy levels and the relative ordering among them. We also look at whether the energy levels coming from the octets are non-degenerate. It is also possible to use these energy levels as input for UChPT and study the trajectories of the poles the starting at the  $SU(3)$ point.

\section{Operator construction}
We construct interpolation operators that belong to the particular $SU(3)$ irreducible representations that we want to study. We start by taking the interpolation operators that belong to the meson and baryon octets, we do the direct product and we find the appropriate linear combinations of these states that transform under the singlet and octets irreducible representations. For each of these representations we then choose the state with zero isospin and minus one strangeness to interpolate the $\Lambda$(1405).

The singlet and the octets have different quantum numbers, so we expect no mixing between them. However, the two octet representations have the same quantum numbers, so there is nothing preventing their mixing. Physically this means that both octet irreducible representations will couple to the same physical states. In UChPT, at the end, the interaction is diagonalized to isolate the physical states \cite{Bruns:2021krp}. In our case, we construct two irreducible representations, $8^{A}$ and $8^{B}$, so that the Wick contractions of states of the two different irreducible representations is zero

\begin{equation}
\contraction[1ex]{\bigl\langle\,}{8^{A}_{i}}{}{\overline{8^{B}_{j}}}
  \left\langle 8^{A}_i  \overline{8^{B}_j} \right\rangle \sim \delta^{AB} \delta_{ij} ,
\end{equation}
with A and B indicating the irreducible representation and $i$ and $j$ indicating the state withing each irreducible representation. Within each irreducible representation, the contraction of any state with itself gives the same result. 

For each irreducible representation we have used a basis of three bilocal interpolation operators
\begin{align*}
\mathcal{O}_1 & \sim \gamma_5 P_{+} q_1 (q_2 C \gamma_5 q_3) (q_4 \gamma_5 \bar{q_5}),\\
\mathcal{O}_2 & \sim \gamma_5 P_{+} \gamma_5 q_1 (q_2 C q_3) (q_4 \gamma_5 \bar{q_5}), \\
\mathcal{O}_3 & \sim  \gamma_5 P_{+} q_1 (q_2 i \gamma_4 C \gamma_5 q_3) (q_4 \gamma_5 \bar{q_5}), \\
\end{align*}
where the meson is interpolated by the pseudoscalar operator with negative parity and zero spin; and the baryon is interpolated by three operators, all with spin one-half and projected into positive parity. The extra $\gamma_5$ is added to ensure the whole operator has negative parity. The operators are calculated at zero total momentum and zero relative momentum between baryon and meson operators.

\section{Calculation details and lattice setup}

Our measurements are performed in gauge configurations generated by the CLS collaboration \cite{Horz:2020zvv,BaryonScattering:2025ziz}. These gauges are generated with a Clover-Wilson action with $N_f = 2+1$ dynamical fermions with $m_u=m_d=m_s$. The lattice spacing is $a\approx 0.086\text{fm}$, $V=48^{3}\times 96$ and $m_\pi L \approx 14.5$, see Table~\ref{table:latt_params} which is very large and allows us to have the exponentially suppressed finite volume effects under control. In total for this work we have used 400 gauge configurations.

\begin{table}[h!]
\begin{center}
	\begin{tabular}{|c c c c c|}
		\hline
		\hline
		Ensemble & Vol & $M_\pi$ & $M_\pi L$ & $N_e$\\
		\hline
		C103 & $48^3\times 96$ & ~714 MeV & ~14.5 & 100\\
		\hline\hline
	\end{tabular}
\end{center}
	\caption{Parameters of the gauge configurations used in the calculations in this work.}
	\label{table:latt_params}
\end{table}

In the calculation of the correlation functions, the presence of quark and antiquark operators of the same flavor in the same timeslice will lead to all to all propagators. For their calculation we use the well established method of distillation \cite{HadronSpectrum:2009krc}. Distillation is a smearing of the quark fields using a truncated space of eigenvectors of the Laplace operator defined on the lattice. The truncation is possible because the higher modes are exponentially suppressed. For the calculations in this work we used a Laplacian subspace spanned by $N_e = 100$ eigenvectors. 

For the UChPT calculations the pion decay constant, $f_\pi$, is needed. Since we are in the $SU(3)$ limit the decay constant will be the same for all pseudoscalar mesons. We have calculated following \cite{Bruno:2016plf} without using the order $a$ improvements. The value obtained is $f_{\text{meson}} = 202.3(9.7)$ MeV.

The statistical uncertainties were estimated using the bootstrapping resampling procedure. The calculation of the eigenvectors and perambulators were performed with PyQuda \cite{Jiang:2024lto}. The calculation of the currents necessary for the calculation of the decay constant were performed with chroma \cite{Edwards:2004sx}. In both cases the inversions were performed using the QUDA library \cite{Clark:2009wm}.

\section{Results}
\begin{figure}[htpb]
  \centering
  \includegraphics[width=0.8\textwidth]{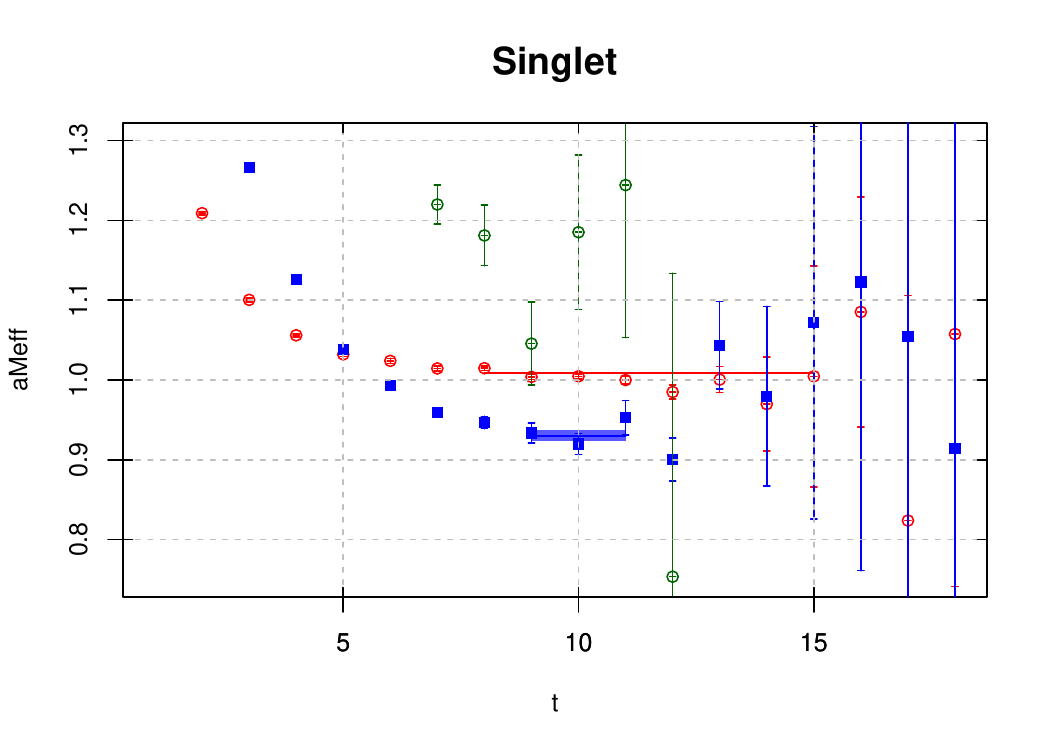}
  \caption{Effective mass of the principal correlators from the GEVP }
  \label{fig:gevp_singlet}
\end{figure}

Here we show the results for the measurements of the energy levels. In Fig.\ref{fig:gevp_singlet} we show the effective mass for the principal correlators of the operator matrix belonging to the singlet representation. We are interested in extracting the ground state, so we look at the lowest energy level which sits below the two particle threshold and is therefore a bound state. 

The first excited state sits at exactly the baryon-meson threshold energy. It is interesting that its effective mass has smaller errors and a better signal than the ground state. The reason is that, because of the baryon-meson structure of the interpolation operators, there is a very large overlap with the baryon-meson states on the lattice, even more than with the lambda state. In fact, the inclusion of the interpolation basis, in particular the operator $\mathcal{O}_2$, is crucial to resolve the bound state.

\begin{figure}[htbp]
\centering
\begin{minipage}{0.48\textwidth}
  \centering
  \includegraphics[width=\linewidth]{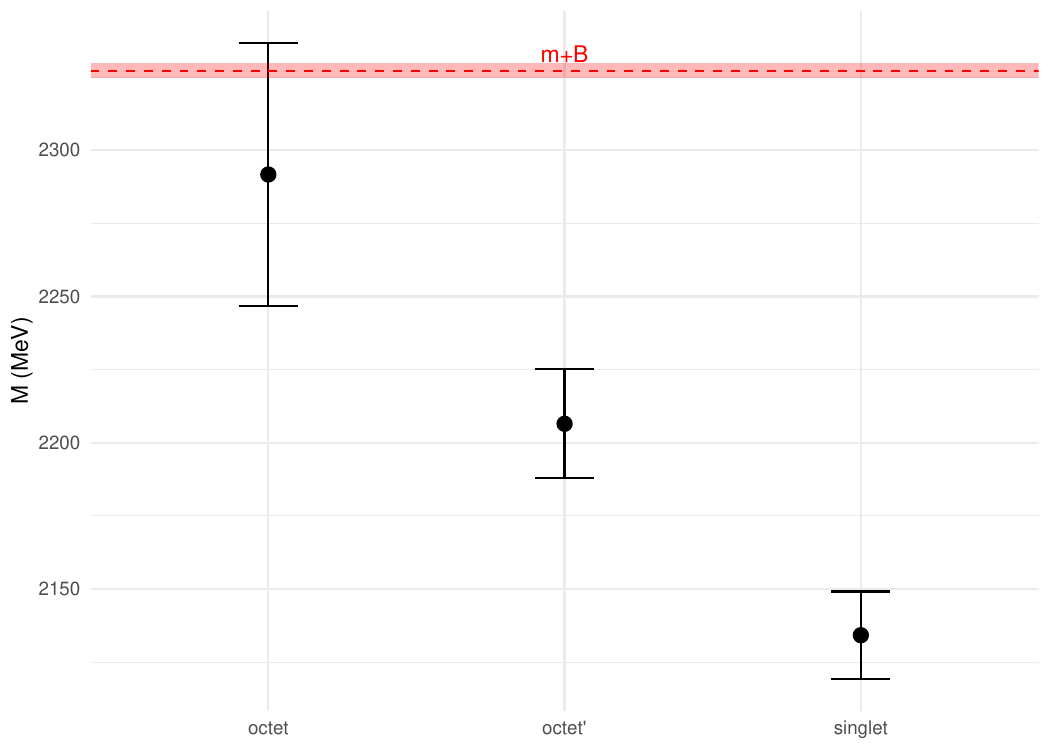}
\end{minipage}\hfill
\begin{minipage}{0.48\textwidth}
  \centering
  \includegraphics[width=\linewidth]{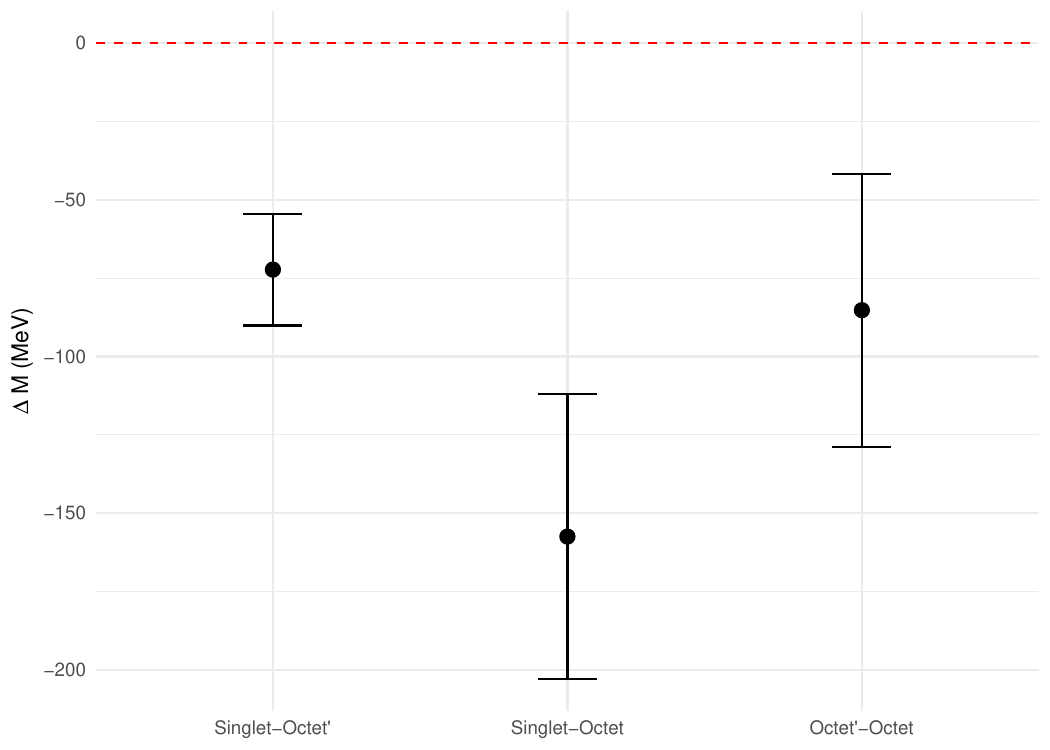}
\end{minipage}
\caption{\textbf{Left}: Energy levels for the singlet, octet and octet prime. The red line indicates the Baryon-Meson threshold. \textbf{Right}: Correlated difference between the energy levels.}
  \label{fig:energy_levels}
\end{figure}

We repeat the same calculation for the correlator matrices of the two octets and we extract the energy levels. The qualitative behavior is similar to the singlet, so we omit these plots and we show directly the ground state energies in Fig. \ref{fig:energy_levels}. From the lattice results we will call $\textbf{8}'$ (octet prime) to the octet state with the lowest energy and $\textbf{8}$ (octet) to the octet state with the highest energy. We found that the energy levels for the singlet and the octet prime are below the baryon meson threshold, so we can consider them bound states. However, the energy level for the octet is compatible with the two particle threshold so we can not say it is a bound state. To elucidate whether it is a bound state or not we will need to increase statistics or enlarge the operator basis. 

Then we look at the ordering of the energy levels and we can see that the singlet is non-degenerate with the octets and that it has lower energy ($E_1 < E_{8}, E_{8'}$). This agrees with previous lattice results \cite{Murakami:2025oig} and with the predictions of UChPT \cite{Oller:2000fj,Jido:2003cb}. Next, we want to look at the splitting between the different states so we perform correlated differences to account for the correlations in the uncertainties. The result is shown in Fig. \ref{fig:energy_levels} on the right plot. We confirm then that the singlet is non degenerate with the octets and we also find that the difference of the two octets is non zero (within one sigma), meaning that they are non degenerate, which agrees with the results from NLO UChPT \cite{Pittler:2025upn}. 

\begin{figure}[htpb]
  \centering
  \includegraphics[width=\textwidth]{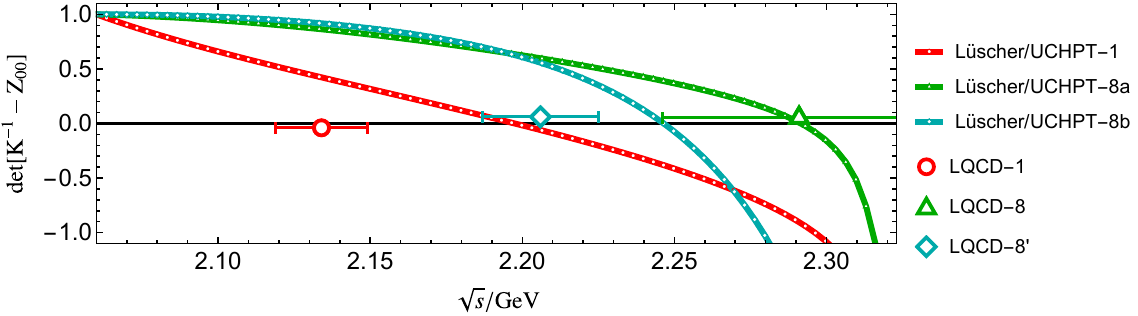}
  \caption{Comparison between the energy levels computed in this work and the predicted by UChPT using \cite{BaryonScatteringBaSc:2023zvt} as input. Figure courtesy of M. Mai.}
  \label{fig:Prediction16}

\end{figure}

As a next step we do a preliminary comparison of our results with a prediction from UChPT. We use the lattice results from \cite{BaryonScatteringBaSc:2023zvt} as input for UChPT \cite{Pittler:2025upn} and we calculate the energy levels at the $SU(3)$ point used in this work. We show the results on Fig. \ref{fig:Prediction16}, there we plot the Lüscher equation projected into the three attractive irreducible representations and diagonalized. The zero crossings are the UChPT predicted energy levels. We refer as $\textbf{8}^A$ to the higher energy level from UChPT, and as $\textbf{8}^B$ to the lowest energy one.  

The points with errors are the energy levels calculated in this work. We find that the level ordering predicted by UChPT agrees with that calculated on the lattice. UChPT also predicts two distinct energy levels coupling to the octet representations and the energy splitting agrees with the one from lattice within error bands. For the exact location of the energy levels we find agreement with the highest energy octet state. For the lower energy octet and the singlet, the UChPT predictions are higher than the calculated lattice results.

\section{Conclusions}

In this work we have measured the energy levels of the singlet and octet irreducible representations at the $SU(3)$ flavor point. To do this we have used Baryon-Meson operators explicitly constructed to belong to the different irreducible representations. We found two bound states one coming from the singlet and one of them coming from one of the octets. The state from the other the two particle threshold. The bound state coming from the singlet is located at a lower energy than the states coming from the two octets. When looking at the two states that come from the octets, a correlated difference indicates that they are two distinct energy levels at one sigma, agreeing with the NLO analysis using UChPT. However, our results are compatible with both states being degenerate within two sigmas, so in the future it is necessary to increase the statistics to confirm this result with greater confidence.

An important step in the future is to increase the operator basis. For this, the inclusion of single baryon operators, that belong to the singlet and octet baryon $SU(3)$ irreducible representations would be the first step. Furthermore, it would be interesting to investigate the impact of 5-quark operators without the clear Baryon-Meson gamma structure (but still belonging to the same irreducible representations) for which the introduction of local operators is needed. Furthermore, it is necessary to have some control over the pion mass dependence of the energy levels calculated above, for which we aim to repeat the calculation in gauges with a pion mass of approximately 450 MeV. This will also allow to have more inputs to constrain the UChPT models.

\section*{Acknowledgements}

We would like to thank A. Walker-Loud for providing us with the gauge configurations used in this work. And  M.Mai for the help with the UChPT part of this work. J.S.S. would also like to thank H. Yan for his help setting up the distillation used in this project.  This work was partly funded by the Deutsche Forschungsgemeinschaft (DFG, German Research Foundation) as part of the CRC 1639 NuMeriQS – Project number 511713970. This work has been supported by the MKW NRW under the funding code NW21-024-A as part of NRW-FAIR.

\end{document}